\documentclass[11pt]{article} 
\usepackage{amsmath,amstext,amsbsy,amssymb}
\usepackage{bm}
\usepackage{cite}
\usepackage{color}
\newcommand{\Tr}{{\rm Tr}\, }

\newcommand{\be}{\begin{equation}}
\newcommand{\ee}{\end{equation}}
\newcommand{\w}{\wedge}

\newcommand{\ssH}{{\scriptscriptstyle{H}}}

\newcommand{\ssh}{{\scriptscriptstyle{1/2}}}
\newcommand{\ssW}{{\scriptscriptstyle{\rm W}}}
\newcommand{\ssA}{{\scriptscriptstyle{\rm A}}}
\newcommand{\ssB}{{\scriptscriptstyle{\rm B}}}
\newcommand{\ssC}{{\scriptscriptstyle{\rm C}}}
\newcommand{\ssD}{{\scriptscriptstyle{\rm D}}}
\newcommand{\ssE}{{\scriptscriptstyle{\rm E}}}

\newcommand{\ssI}{{\scriptscriptstyle{\rm I}}}
\newcommand{\ssJ}{{\scriptscriptstyle{\rm J}}}
\newcommand{\ssK}{{\scriptscriptstyle{\rm K}}}
\newcommand{\ssL}{{\scriptscriptstyle{\rm L}}}
\newcommand{\ssM}{{\scriptscriptstyle{\rm M}}}

\newcommand{\ssV}{{\scriptscriptstyle{V}}}

\long\def\symbolfootnote[#1]#2{\begingroup%
\def\thefootnote{\fnsymbol{footnote}}\footnote[#1]{#2}\endgroup}

\begin{document}

\begin{center}

{\Large \bf   

\vspace{2cm}

A Semiclassical Formulation  of the Chiral Magnetic Effect and Chiral Anomaly in Even $d+1$  Dimensions

}

\vspace{2cm}

\"{O}mer F. Dayi$^*$ and Mahmut Elbistan$^{*,\dagger}$

\vspace{5mm}

{\em {\it ${}^*$Physics Engineering Department, Faculty of Science and
Letters,\\ Istanbul Technical University,
TR-34469, Maslak--Istanbul, Turkey\\
${}^\dagger$ Institute of Modern Physics, Chinese Academy of Sciences, Lanzhou, China}}
\footnote{{\it E-mail addresses:} dayi@itu.edu.tr , elbistan@impcas.ac.cn}

\end{center}

\vspace{3cm}

In terms of the matrix valued Berry gauge field strength   for the
Weyl Hamiltonian in any even spacetime dimensions a symplectic form whose elements are matrices in spin indices is introduced.
 Definition of the volume form is  modified appropriately. A simple method of finding the path integral measure and the chiral current
in the presence of external electromagnetic fields is presented. 
It is shown that 
within this new approach  the chiral magnetic effect as well as  the  chiral anomaly in even $d+1$ dimensions are accomplished 
straightforwardly. 

\vspace{2cm}

\newpage

\section{Introduction}

Semiclassical analysis of dynamical systems naturally possesses  ambiguities:  It is defined within a classical phase space, though it claims to 
embrace  quantum mechanical effects. Nevertheless, it is useful to get a better understanding of some quantum mechanical phenomena 
especially in many body systems. In this respect incorporating quantum mechanical effects in classical kinetic theory 
is of concern. An outstanding achievement for this purpose was to embody chiral anomaly in classical kinetic theory \cite{soy,sy}.
In \cite{soy}  $3+1$ spacetime dimensional Fermi liquid theory was modified introducing the  Berry curvature which is a monopole field deforming the initial phase space \cite{dhh}. They successfully obtained the chiral magnetic effect and chiral anomaly in external electromagnetic fields.  
The same results were accomplished in \cite{sy} by  showing that  the monopole field is the field strength of the Berry gauge field emerging from the Weyl Hamiltonian of a massless Dirac particle in $3+1$ dimensions. Recently, several aspects of the chiral kinetic theory were  studied extensively \cite{ckt, dtson, gbasar, dcapasso, huyee, kland, dekhar}.  However, in higher dimensions calculation of the chiral magnetic effect and chiral anomaly within the same formulation is missing. In fact the chiral magnetic effect and the chiral vortical effect for even $d+1$ dimensions were conjectured \cite{ls}. These conjectures were supported with the results of \cite{log}. Our main goal is to calculate the chiral magnetic effect and chiral anomaly in even $d+1$ dimensions within the same theory. We will show that the source of both effects is the Dirac monopole field defined through the Berry field strength.

Incorporating spin in classical description of particles is one of the intriguing questions of semiclassical dynamics. For instance in spacetime dimensions higher than $3+1,$ 
 Weyl Hamiltonians
lead to the  non-Abelian Berry gauge fields which are matrices in ``spin indices".  A formulation of  anomalies in  higher dimensional classical chiral kinetic theory   was carried out in \cite{ds}. They surmount the difficulty of treating the matrix valued Berry gauge fields by introducing group valued degrees of freedom and ``dequantizing spin". However this makes unclear the explicit form of the equations of motion which would yield a  generalization of the chiral magnetic effect in higher dimensions. On the other hand, it is possible to deal with matrix valued Hamiltonians in the presence of the Berry gauge fields systematically \cite{ofd}. Inspired with this approach, though following mainly the differential form methods employed in \cite{ds}, we present a formalism of the classical chiral  kinetic theory in even $d+1$ dimensions yielding  both the  chiral  anomaly and chiral magnetic effect in external electromagnetic fields. We also show that the chiral vortical effec!
 t can be formulated similarly.

The main ingredient of our approach is to extend the Hamiltonian methods  to  cover the symplectic  forms which are matrix valued
in spin space. Considering $d+1$ spacetime dimensional systems,  elements of the   symplectic matrix are labeled as usual with the phase space indices $a,b=1,\cdots, 2d.$  But, for $d>3$ we let them to be matrices in  spin space. Therefore, the symplectic matrix elements carry   the phase space  indices $a,b,$ as well as  the spin  indices  $\alpha,\beta=1,\cdots, 2^{[(d-3)/2]}, $  but we distinguish between  these indices.
The former ones label the dynamical variables of the semiclassical  kinetic theory  where probability functions may depend only on classical phase space variables. Hence,  we  define the equations of motion as matrix valued in spin space. This is in accord with the fact that
in a Fermi liquid theory the quasi-particles are described in general with matrices in the spin indices \cite{pl}. However, 
it is usually sufficient to consider distributions yielding scalar particle density.  Thus a procedure of reducing our matrix valued quantities to scalars should be given. First of all,
we need an appropriate definition of   measure for the phase space path integral which should  be a scalar. The most orthodox choice would be to define it as the determinant of the related  
$(2d\ 2^{[(d-3)/2]}) \times (2d\ 2^{[(d-3)/2]})$ dimensional  matrix, considering  phase space and spin indices on the same footing. However, we distinguish between the phase space and spin indices. Thus, we first ignore the spin indices and obtain the Pfaffian of  symplectic matrix considering only  the phase space  indices $a,b.$ This Pfaffian will be a matrix in  the spin indices $\alpha,\beta.$ We define the measure as  its trace. Similarly, to define the currents within the kinetic theory we take the trace of the related matrix valued currents accompanied by the reduction of the matrix valued velocities to scalars. We will demonstrate  that our method permits us to calculate  the $d+1$ dimensional   chiral magnetic effect.
It coincides with the conjectured chiral magnetic effect in \cite{ls}. We will also comment that the conjectured chiral vortical effect can be derived similarly. Moreover, within the same formalism the chiral anomaly in even $d+1$ dimensions is engendered correctly.

In Section \ref{section2} we recall the definition of the Berry gauge fields for  Weyl  Hamiltonians in general. 
In Section \ref{section3}
the $3+1$ dimensional semiclassical chiral kinetic theory is considered using the differential form formalism of dynamical systems. We basically review the 
approach  of \cite{ds} to the chiral (non-Abelian) anomaly though we deal only with the external electromagnetic fields. However, we also show  how this  formalism can be used to achieve
the phase space measure as well as the solutions of the equations of motion for the first time  derivatives  of the phase space variables straightforwardly.

The main results are presented in  Section \ref{section5}.  We present
a method of  extending  the formulation of  Section \ref{section3} to cover the  symplectic forms which are matrices in spin indices. 
We demonstrate that
it permits us to find not only the chiral  anomaly but also the $d+1$ dimensional chiral magnetic effect as conjectured in \cite{ls}. We show how our method can be used to obtain the conjectured chiral vortical effect.

In Section \ref{sectionD} we outlined  our results and  discussed its applicability  to  some other physical system.

\renewcommand{\theequation}{\thesection.\arabic{equation}}
\setcounter{equation}{0}

\section{Weyl Hamiltonian and the Berry Gauge Field
\label{section2}}

  The massless Dirac Hamiltonian  in $d+1$ dimensions  is 
\begin{equation}
\label{dh}
{\cal{H}}=\bm{\alpha}\cdot{\bm{p}},
\end{equation}
in  $\hbar=c=1,$ units. 
$\bm p$ is  the $d$ dimensional momentum vector  and  $\alpha_\ssA;\  {\rm{A}}=1,...,d,$  are  the  $2^{[(d+1)/2]}\times 2^{[(d+1)/2]}$ dimensional matrices  which satisfy the anticommutation relations
\begin{equation}
\{\alpha_\ssA,\alpha_\ssB\}=2\delta_{\ssA\ssB}.
\label{alf}
\end{equation}
They can be expressed as $\bm{\alpha}=\gamma^0 \bm{\gamma},$ in terms of 
the gamma matrices   which obey the Clifford algebra,
$
\{\gamma^{\mu},\gamma^{\nu}\}=2g^{\mu\nu},
$
 where the metric tensor is $g^{\mu\nu}= {\rm diag } (1,-1,\cdots,-1);\ \mu, \nu =0,1,...,d.$   

In the  chiral representation  $\bm\alpha $ is block diagonal, so that (\ref{dh}) yields the Weyl Hamiltonian
\begin{equation}
{\cal{H}}_\ssW=\bm{\Sigma}\cdot \bm{p},
\label{WHa}
\end{equation}
where $\Sigma_\ssA$ are $2^{[(d-1)/2]}\times 2^{[(d-1)/2]}$ matrices.
Quantum mechanical description  of the Weyl particle is furnished with 
the  eigenvalue equation
$$
{\cal{H}}_\ssW \psi_E (\bm p )= E\psi_E (\bm p ).
$$
The energy eigenvalues are $E=(p,-p )$ where $p=|\bm p|$.
Focusing on the positive energy solutions $|\psi^{\alpha}\rangle ;\ \alpha =1,...,2^{[\frac{d-3}{2}]},$ one can define the Berry gauge field as
\begin{equation}
{\cal{A}}_\ssA^{\alpha\beta}=i \langle \psi^{\alpha}| \frac{\partial}{\partial p_\ssA} |\psi^{\beta} \rangle .
\label{bgf}
\end{equation}
Although ${\cal{A}}$ is Abelian for $d=3,$ it  becomes to be  non-Abelian for higher dimensions when there is degeneracy. Thus,  in general the Berry field strength is given by
\begin{equation}
{\cal{G}}_{\ssA\ssB}^{\alpha\beta}=\frac{\partial {\cal{A}}_\ssB^{\alpha\beta}}{\partial p_\ssA} -\frac{\partial {\cal {A}}_\ssA^{\alpha\beta}}{\partial p_\ssB} -i[{\cal{A}}_\ssA,{\cal{A}}_\ssB]^{\alpha\beta}.
\label{dbc}
\end{equation}

  We will  present the $d=3$ and $d=5$ Berry gauge fields explicitly, respectively, in the next section and Appendix \ref{section4}.

\renewcommand{\theequation}{\thesection.\arabic{equation}}
\setcounter{equation}{0}

\section{The Chiral Anomaly and Chiral Magnetic Effect in $\bm{3+1}$ Dimensions}   
\label{section3}

In \cite{sy}  it was observed that for $d=3,$  the Berry field strength  extracted from the diagonalization of Weyl Hamiltonian describes
a monopole located  at  $\bm p =0.$    They clarified the role of
 this monopole field  in obtaining the CME and the chiral anomaly.  In \cite{ds}  it was argued that in all even dimensions chiral  (non-Abelian) anomaly arises in the presence of the Berry gauge fields within the differential form
 formulation of classical Hamiltonian dynamics.  Here we  basically review the approach of \cite{ds}  considering  external electromagnetic fields in $d=3.$ However, we will also show how the differential form formalism can be employed to acquire the first time derivatives of phase space variables which are needed to formulate  the CME as in \cite{sy}.

In the chiral representation of gamma matrices
\be
\label{cb}
\gamma^0=\begin{pmatrix} 0 & -1\\ -1 & 0 \end{pmatrix}, 
\bm{\gamma}= \begin{pmatrix} 0 & \bm{\sigma}\\ -\bm{\sigma} & 0 \end{pmatrix},
\gamma^5=\begin{pmatrix} 1 & 0\\ 0 & -1 \end{pmatrix},  
\ee
where $\bm\sigma=(\sigma_1, \sigma_2,\sigma_3)$ are the Pauli spin matrices, the
Weyl Hamiltonian is acquired: 
\be
\label {wh3}
{\cal{H}_\ssW}^{(3)}=\bm{\sigma}\cdot\bm{p}=\begin{pmatrix}p_3 & p_1-ip_2\\ p_1+ip_2 & -p_3 \end{pmatrix}.
\ee

To formulate the Hamiltonian dynamics  one introduces  the one-form
$$
\eta_{\ssH}=p_a dx_a+A_a(x,t)dx_a-{\cal{A}}_a(p) dp_a-H dt,
$$
where $a=1,2,3.$  We work in $e=c=1,$ units. $\bm{A}$ is the  vector potential of  the external  magnetic field
$\bm{B}:$
$$
F_{ab}=\frac{\partial A_b}{\partial x_a}-\frac{\partial A_a}{\partial x_b}=\epsilon_{abc}B_c.
$$
 $H=p+A_0$ where $p$ is the Weyl Hamiltonian (\ref{wh3})  diagonalized and projected onto the positive energy eigenstate. $A_0(x,t)$ is the scalar potential of the external electric field  $\bm{E}=\partial \bm A/ \partial t-\bm{\nabla}A_0.$ ${\cal{A}}_a (p)$ denotes  the   Abelian Berry gauge field.
Inspecting the positive energy solution of  (\ref{wh3}) one can show that its  field strength is given in terms of $\bm{b}=\bm{\hat{p}}/2p^2,$ as
\be
\label{bg3}
G_{ab}=\frac{\partial {\cal{A}}_b}{\partial p_a}-\frac{\partial {\cal{A}}_a}{\partial p_b}=\epsilon_{abc}b_c.
\ee    
It is the field of a monopole located at $\bm{p}=0:$  $\bm{\nabla}\cdot \bm{b}=2\pi\delta^3(p).$

 The exterior derivative of $\eta_{\ssH}$ provides   the symplectic  two-form  
$$
w_{\ssH}=d\eta_{\ssH} = dp_a\wedge dx_a+F -G-\hat{p}_adp_a\wedge dt+E_adx_a\wedge dt,
$$
where
$F=\frac{1}{2} F_{ab}\ dx_a\wedge dx_b$ and $G=\frac{1}{2}G_{ab}\ dp_a\wedge dp_b.$
 In terms of $w_{\ssH}$ the volume form in $6+1$ dimensional phase space   is defined by
\be
\label{vform3}
\Omega \equiv \frac{1}{3!}  w_\ssH^3 \w dt.
\ee
It can be written in terms of the  canonical volume element  of the phase space $dV^{(3)}$  as
\be
\label{vform3l}
\Omega =\sqrt{w}\ dV^{(3)}\w dt .
\ee
Here,
$\sqrt{w}\equiv\sqrt{\det (w)},$  is the Pfaffian of the matrix
$$
\begin{pmatrix}
{F}_{ab} &-\delta_{ab}\\
\delta_{ab} & -G_{ab}
\end{pmatrix} .
$$
It was  calculated explicitly in \cite{dhh} as $\sqrt{w}=(1+\bm{B}\cdot\bm{b})$.

The  equations of motion can be derived by demanding that  $w_\ssH$ satisfies
\begin{equation}
i_v w_\ssH=0,
\label{int0}
\end{equation}
which  is the interior product   of the vector field
$$
v=\frac{\partial}{\partial t}+\dot{x}_a\frac{\partial}{\partial x_a}+\dot{p}_a\frac{\partial}{\partial p_a} ,
$$
with $w_\ssH.$
Indeed, (\ref{int0}) gives rise to the equations of motion obtained in \cite{sy}:
\begin{subequations}
\label{eom3}
\begin{align}
\dot{p}_a=\dot{x}_b{F}_{ab}+E_a,\\
\dot{x}_a=\dot{p}_bG_{ab}+\hat{p}_a.
\end{align}
\end{subequations}

Liouville equation will be provided by the time evolution of the volume form $\Omega$ which can be found by calculating its
 Lie derivative associated with $v.$ 
It can be  accomplished in two different ways. First  one can employ (\ref{vform3l}) to observe that
\be
\label{le3}
L_v \Omega=(i_v d+d i_v)\left(\sqrt{w}dV^{(3)}\w dt \right) =\left (\frac{\partial}{\partial t}  \sqrt{w}+\frac{\partial}{\partial x_a}(\sqrt{w} \dot{x}_a)+\frac{\partial}{\partial p_a} (\sqrt{w} \dot{p}_a )\right )dV^{(3)}\w dt.
\ee
Then, one can show that the definition  (\ref{vform3})   implies
$$
L_v\Omega=
(i_vd+di_v)\left(\frac{1}{3!} w_\ssH^3 \w dt\right)
=di_v\left(\frac{1}{3!} w_\ssH^3 \w dt\right)=
\frac{1}{2}dw_\ssH\wedge w_\ssH^2,
$$
where
$$
dw_\ssH=-\frac{1}{2}\frac{\partial G_{ab}}{\partial p_c}dp_c\w dp_a\w dp_b.
$$ 
Making use of (\ref{bg3}) one can express $dw_\ssH$ as
$$
dw_\ssH=-(\bm{\nabla}\cdot\bm{b})\ dp_1\w dp_2\w dp_3=-2\pi\delta^3(p)\ dp_1\w dp_2\w dp_3.
$$
The unique contribution  from $w_{\ssH}^2$ will be $E_aF_{bc} dx_a\w  dx_b\w  dx_c\w dt,$ thus one finds that
$$
\frac{1}{2}dw_\ssH\wedge w_\ssH^2=2\pi\delta^3(p)(\bm{E}\cdot \bm{B})\ dV^{(3)}\w dt.
$$
One reaches the conclusion that  the Liouville equation is anomalous:
\be
\label{aterm3}
\left (\frac{\partial}{\partial t} \sqrt{w}+\frac{\partial}{\partial x_a}(\sqrt{w}\dot{x}_a)+\frac{\partial}{\partial p_a} (\sqrt{w} \dot{p}_a )\right )=2\pi\delta^3(p)\bm{E}\cdot\bm{B}.
\ee

In quantum field theory the chiral anomaly is expressed as  non-conservation of the classically conserved chiral current at the quantum level. In order to connect the anomaly contribution  in (\ref{aterm3}) to  the non-conservation of particle number   introduce the probability function $ f(x,p,t),$ which satisfies the collisionless Boltzmann equation
\be
\label{cbe3}
\frac {df}{dt}=\frac{\partial f}{\partial t}+\frac{\partial f}{\partial x_a}\dot{x}_a+\frac{\partial f}{\partial p_a}\dot{p}_a=0.
\ee     
It is appropriate to define the probability density function as
$
\rho(x,p,t)=\sqrt{w}f.
$
Hence, define the particle number density and the particle current density:
$$
n(x,t)=\int \frac{d^3p}{(2\pi)^3}\rho (x,p,t),\ \ \
j_a=\int \frac{d^3p}{(2\pi)^3}\rho (x,p,t) \dot{x}_a.
$$
Utilizing (\ref{aterm3}) and (\ref{cbe3}) one can observe that
\be
\label{ncp3}
\frac{\partial \rho}{\partial t}+\frac{\partial (\rho \dot{x}_a)}{\partial x_a}+\frac{\partial (\rho \dot{p}_a)}{\partial p_a}=2\pi f\delta^3(p)\bm{E}\cdot\bm{B}.
\ee  
 In order to derive  non-conservation of the particle current integrate (\ref{ncp3})  over  momentum degrees of freedom.
There is no influx of particles from the negative energy sea as we only deal  with the positive energy sector of the Weyl Hamiltonian. 
Thus, 
the momentum  current density $\bm{j}_p=\rho\bm{\dot{p}}$ vanishes at the boundary of the momentum space:
$$
 \int \frac{d^3p}{(2\pi)^3} \frac{\partial (\rho \dot{p}_a )}{\partial p_a}=0.
$$
Actually, the Berry monopole situated  on the boundary $|\bm{p}|=0$ is responsible for the non-conservation of the chiral particle current. A detailed  explanation was given in \cite{sy}. Therefore, the statement of
 non-conservation of the particle current follows, 
$$
\frac{\partial n(x,t)}{\partial t}+\bm{\nabla} \cdot \bm{j}=\frac{1}{4\pi^2}f(x,p=0,t)\bm{E}\cdot\bm{B}.
$$

To derive  the CME in line with \cite{sy} one needs to solve the equations of motion  (\ref{eom3}) for $\sqrt{w}\dot{x}_a.$  Obviously, here this can be  done  directly, because  $\sqrt{w}$ has already been calculated in \cite{dhh}. 
However, it is possible to acquire  the same  result  by  inspecting the explicit form of $w_\ssH^3.$ 
As a byproduct the explicit form of 
$\sqrt{w}$ will also be provided. To
 present this method which will be  extremely useful in higher dimensions, 
 let us write $w_\ssH^3$ explicitly as 
\begin{eqnarray*}
w_\ssH^3&=&dp_a\w dx_a\w dp_b\w dx_b\w dp_c\w dx_c-6{F}\w {\cal{G}} \w dp_a\w dx_a
\\
&&-3\hat{p}_a dp_a\w dt\w dp_b\w dx_b\w dp_c\w dx_c -6E_a {F}\w {\cal{G}}\w dx_a\w dt
\nonumber\\
&&+3E_a dx_a\w dt\w dp_b\w dx_b\w dp_c\w dx_c-6\hat{p}_a{F}\w dp_a\w dt\w dp_b\w dx_b\nonumber \\
&&-6E_a{\cal{G}}\w dx_a\w dt\w dp_b\w dx_b +6\hat{p}_a{F}\w {\cal{G}}\w dp_a\w dt. 
\end{eqnarray*}
Its Lie derivative associated with $v$  procures
\begin{eqnarray}
\label{le3ex} 
L_v\Omega=\frac{1}{3!}dw_\ssH^3&=& \Big\{\frac{\partial}{\partial t}(1+\bm{B}\cdot\bm{b})+\frac{\partial}{\partial\bm{x}}\left(\bm{\hat{p}}+\bm{E}\times\bm{b}+\bm{B}(\bm{\hat{p}}\cdot \bm{b})\right) \nonumber \\
&&+\frac{\partial}{\partial \bm{p}}\left(\bm{E}+\bm{\hat{p}}\times\bm{B}+\bm{b}(\bm{E}\cdot \bm{B})\right) \Big\} \ 
dV^{(3)}\w dt.
\end{eqnarray} 
Comparing (\ref{le3ex})  with (\ref{le3}) one can directly read  $\sqrt{w},$ as well as the solutions of the equations of motion (\ref{eom3}) as 
\begin{eqnarray*}
\sqrt{w}&=&1+\bm{B}\cdot\bm{b},\nonumber\\
\sqrt{w}\bm{\dot{x}}&=&\bm{\hat{p}}+\bm{E}\times\bm{b}+\bm{B}(\bm{\hat{p}}\cdot \bm{b}),\\
\sqrt{w}\bm{\dot{p}}&=&\bm{E}+\bm{\hat{p}}\times\bm{B}+\bm{b}(\bm{E}\cdot \bm{B}) .\nonumber
\end{eqnarray*} 
Now,  the particle current density $\bm{j}$ can be obtained as 
$$
\bm{j}=\int \frac{d^3p}{(2\pi)^3} \sqrt{w}\bm{\dot{x}}f=\int \frac{d^3p}{(2\pi)^3} \bm{\hat{p}}f+\bm{E}\times\int \frac{d^3p}{(2\pi)^3}\bm{b}f+\bm{B}\int \frac{d^3p}{(2\pi)^3}\bm{\hat{p}\cdot b}f.
$$
The last term where the current is parallel to the magnetic field is the CME term that was mentioned in \cite{sy}.

\renewcommand{\theequation}{\thesection.\arabic{equation}}
\setcounter{equation}{0}

\section{Chiral Kinetic Theory in $d+1$ Dimensions
\label{section5}}

We would like to consider  kinetic theory of the $d+1=2n+2;\ n=1,2...,$ dimensional Weyl particles  in the presence of the
Berry gauge field (\ref{bgf}), the external magnetic  field 
${\cal{F}}_{\ssA \ssB}=\partial A_\ssB/ \partial x_\ssA-\partial A_\ssA/ \partial x_\ssB,$ and the electric field ${\cal{E}}_\ssA=-\partial A_0/ \partial x_\ssA+\partial A_\ssA/ \partial t,$ pointing towards the ${\hat{x}}_\ssA$ direction.
 $A_0 (x,t)$ and $\bm{A} (x,t),$  are the scalar and vector potentials.
We will mainly follow the procedure of Section \ref{section3} with some modifications  required
to deal with spin degrees of freedom.  

Dealing with the non-Abelian  fields ${\cal{A}}_\ssA,$ the  appropriate definition of the matrix valued symplectic two-form is
$$
\tilde{W}_\ssH\equiv dp_\ssA\wedge dx_\ssA+{\cal{F}}-{\cal{G}}-\hat{p}_\ssA dp_\ssA\wedge dt+{\cal{E}}_\ssA dx_\ssA \wedge dt,
$$
where, ${\cal{F}}=\frac{1}{2}{\cal{F}}_{\ssA\ssB}dx_\ssA\w dx_\ssB,$ and ${\cal{G}}=\frac{1}{2}{\cal{G}}_{\ssA\ssB}dp_\ssA\w dp_\ssB .$
The symplectic two-form $\tilde{W}_\ssH$ is a $2^{n-1}\times 2^{n-1}$ matrix in spin indices.
We suppress the matrix indices and do not write the  unit matrix $I$ explicitly, unless it is necessary. 

Let the  matrix valued $(\dot{X}_\ssA , \dot{P}_\ssA)$ denote  time evolution of the phase space variables $(x_\ssA, p_\ssA )$ which are acquired  through the interior product of the matrix valued vector field 
$$
\tilde{V}=\frac{\partial}{\partial t}+\dot{X}_\ssA \frac{\partial}{\partial x_\ssA}+\dot{P}_\ssA \frac{\partial}{\partial p_\ssA}, 
$$
with the symplectic two-form $\tilde{W}_\ssH$ as
$$
i_{\tilde{\ssV}}\tilde{W}_\ssH=0.
$$
This gives rise to
\begin{subequations}
\label{eomd}
\begin{align}
\dot{P}_\ssA=\dot{X}_\ssB{\cal{F}}_{\ssA\ssB}+{\cal{E}}_\ssA, \\
\dot{X}_\ssA={\cal{G}}_{\ssA\ssB}\dot{P}_\ssB+\hat{p}_\ssA ,
\end{align}
\end{subequations}
as the equations of motion. We would like to stress that $(\dot{X}_\ssA , \dot{P}_\ssA)$ are matrices.

\subsection{Liouville Equation and the Chiral Anomaly}

The novelty in this formulation is the fact that we do not treat spin degrees of freedom  as  dynamical variables. Classical dynamics is asserted through  the phase space variables $(\bm x , \bm p ),$ so that  we   
define  the phase space volume  form as
\be
\label{vformd}
\tilde{\Omega}\equiv\frac{(-1)^{n+1}}{(2n+1)!}\tilde{W}_\ssH^{2n+1}\w dt.
\ee
We  express it in terms of the $2d$ dimensional  Liouville measure $dV,$ as   
\be
\label{vformdl}
\tilde{\Omega}= \tilde{W}_\ssh dV\w dt. 
\ee
$\tilde{W}_\ssH$ is a two-form in the phase space variables $(x_\ssA,p_\ssA),$ so that
$\tilde{W}_\ssh$ is the Pfaffian of the $(4n+2)\times(4n+2)$ matrix
$$
\begin{pmatrix}
{\cal{F}}_{\ssA\ssB} & -\delta_{\ssA\ssB}\\
\delta_{\ssA\ssB} & -{\cal{G}}_{\ssA\ssB}
\end{pmatrix}.
$$

In order to derive the Liouville equation we need the 
Lie derivative of the volume form $\tilde{\Omega}.$ By making use of  (\ref{vformdl}) it can be written formally as
\be
\label{led}
L_{\tilde{\ssV}} \tilde{\Omega}=(i_{\tilde{\ssV}} d+d i_{\tilde{\ssV}})\tilde{W}_\ssh \  dV\w dt=\left (\frac{\partial}{\partial t} \tilde{W}_\ssh+\frac{\partial}{\partial x_\ssA}(\dot{X}_\ssA\tilde{W}_\ssh )+\frac{\partial}{\partial p_\ssA} (\tilde{W}_\ssh \dot{P}_\ssA)\right )dV\w dt.
\ee 
However, in order to acquire  it explicitly the definition (\ref{vformd}) should be employed,
\begin{equation}
L_{\tilde{\ssV}} \tilde{\Omega}=\frac{(-1)^{n+1}}{(2n+1)!}d\tilde{W}_\ssH^{2n+1}.
\label{ln1}
\end{equation}
Among the several terms in $\tilde{W}_\ssH^{2n+1}$, the chiral anomaly stems from the one which includes the singularity in momentum space:
\begin{equation}
\frac{(-1)^n(2n+1)!}{(n!)^2}\overbrace{{\cal{G}}...{\cal{G}}}^{n\ times}{\cal{E}}_\ssA\overbrace{{\cal{F}}...{\cal{F}}}^{n\ times}dx_{\ssA}\w dt.
\label{ate}
\end{equation} 
We suppress wedge products of the curvature forms.

To describe quantum mechanical particles possessing spin within classical phase space we  introduced matrix valued quantities representing the spin degrees of freedom. On the other hand,   measure of the related path integral  should be a scalar.
Thus an appropriate definition of the  classical limit  is needed.
For accomplishing the semiclassical approximation we take  the trace over  spin indices and adopt  $\Tr \tilde{W}_\ssh $ as  the  definition of the related path integral measure. 
 As it is demonstrated in Appendix B after taking the trace the unique term which survives in (\ref{ln1}) is originated from (\ref{ate}) and it leads to 
the following anomalous Liouville equation,
$$
\Tr [L_{\tilde{\ssV}} \tilde{\Omega}]=-\frac{1}{(n!)^2}\Tr [d(\overbrace{{\cal{G}}...{\cal{G}}}^{n\ times})]{\cal{E}}_\ssA\overbrace{{\cal{F}}...{\cal{F}}}^{n\ times}dx_{\ssA}\w dt.
$$
In Appendix B  we also briefly reported, following  \cite{me}, how  the singularity is calculated to be   
\begin{eqnarray*}
\Tr [d(\overbrace{{\cal{G}}...{\cal{G}}}^{n\ times})]&=&\frac{1}{2^n}\frac{\partial}{\partial p_{\ssA}}\Tr[\epsilon_{\ssA\ssB\ssC...\ssD\ssE}\overbrace{{\cal{G}}_{\ssB\ssC}...{\cal{G}}_{\ssD\ssE}}^{n\ times}]d^{2n+1}p\nonumber \\
&=&\frac{(-1)^{n+1}(2n)!}{2^n}(\bm{\nabla}\cdot\bm{b})d^{2n+1}p ,
\end{eqnarray*}
where $\bm{b}$ is the $2n+1$ dimensional Dirac monopole field 
\begin{equation}
\bm{b}=\frac{\bm{p}}{2p^{2n+1}}.
\label{dmd}
\end{equation}
Obviously it satisfies
$$
  \bm{\nabla}\cdot\bm{b}=\frac{{\rm{Vol}}(S^{2n})}{2}\delta^{2n+1}(p),
$$
where ${\rm{Vol}}(S^{2n})$ denotes  volume of the $2n$-sphere. Hence, we attain 
\be
\label{ledr}
\Tr[L_{\tilde{\ssV}}\tilde{\Omega}]=\frac{(-1)^{n+1}(2n)!}{(n!)^2\  2^{2n+1}}{\rm{Vol}}(S^{2n})\delta^{2n+1}(p)\epsilon_{\ssA\ssB\ssC...\ssD\ssE}{\cal{E}}_\ssA\overbrace{{\cal{F}}_{\ssB\ssC}...{\cal{F}}_{\ssD\ssE}}^{n\ times}dV\w dt.
\ee 
On the other hand,
within the semiclassical approximation we define 
$$
\sqrt{W}\equiv\Tr[\tilde{W}_\ssh],\  \Tr[\dot{X}_\ssA \tilde{W}_\ssh ]\equiv\sqrt{W}\dot{x}_\ssA,\  \Tr [\tilde{W}_\ssh {\dot{P}_\ssA} ]
\equiv\sqrt{W}\dot{p}_\ssA,
$$
where $(\dot{x}_\ssA ,\dot{p}_\ssA)$ denote the classical velocities. Hence, equating the trace of (\ref{led})  with (\ref{ledr}) we obtain the semiclassical anomalous Liouville equation as
$$
\left (\frac{\partial}{\partial t} \sqrt{W}+\frac{\partial}{\partial x_\ssA}(\sqrt{W}\dot{x}_\ssA )+\frac{\partial}{\partial p_\ssA} (\sqrt{W} \dot{p}_\ssA)\right )=\frac{(-1)^{n+1}(2n)!}{(n!)^2\  2^{2n+1}}{\rm{Vol}}(S^{2n})\delta^{2n+1}(p)\epsilon_{\ssA\ssB\ssC...\ssD\ssE}{\cal{E}}_\ssA\overbrace{{\cal{F}}_{\ssB\ssC}...{\cal{F}}_{\ssD\ssE}}^{n\ times}.
$$
In order to connect it to the non-conservation of the chiral particle number we employ the phase space distribution $f(x,p,t)$ satisfying the collisionless Boltzmann equation 
$$
\frac{\partial f}{\partial t}+\frac{\partial f}{\partial x_\ssA}\dot{x}_{\ssA}+\frac{\partial f}{\partial p_\ssA}\dot{p}_{\ssA}=0,
$$
and define the probability density by $\rho(x,p,t)=\sqrt{W}f$. Thus we find
\begin{equation}
\frac{\partial \rho}{\partial t}+\frac{\partial (\rho \dot{x}_\ssA)}{\partial x_\ssA}+\frac{\partial (\rho \dot{p}_\ssA)}{\partial p_\ssA}=\frac{(-1)^{n+1}(2n)!}{(n)!^2\ 2^{2n+1}}{\rm{Vol}}(S^{2n})f\delta^{2n+1}(p)
\epsilon_{\ssA\ssB\ssC...\ssD\ssE}{\cal{E}}_\ssA\overbrace{{\cal{F}}_{\ssB\ssC}...{\cal{F}}_{\ssD\ssE}}^{n\ times} ,
\label{ncr}
\end{equation}
which is the non-conservation of the phase space probability. Now,  introduce the chiral particle density $n(x,t)=\int{\frac{d^{2n+1} p}{(2\pi)^{2n+1}}\rho}$ and the chiral current density $j_\ssA=\int{\frac{d^{2n+1} p}{(2\pi)^{2n+1}}\rho\dot{x}_\ssA},$ for establishing   non-conservation of the chiral current as
\begin{equation}
\frac{\partial n}{\partial t}+\vec{\nabla} \cdot \vec{j}=\frac{(-1)^{n+1}}{n!\ 2^{n}\ (2\pi)^{n+1}}f(x,p=0,t) \epsilon_{\ssA\ssB\ssC...\ssD\ssE}{\cal{E}}_\ssA\overbrace{{\cal{F}}_{\ssB\ssC}...{\cal{F}}_{\ssD\ssE}}^{n\ times}.
\label{dane}
\end{equation}
It is derived by 
integrating  (\ref{ncr}) over the momentum space and setting ${\rm{Vol}}(S^{2n})=\frac{2^{2n+1}\pi^nn!}{(2n)!}$. Obviously, we also employed
$
\int \frac{d^{2n+1} p}{(2\pi)^{2n+1} } \frac{\partial (\rho \dot{p}_i)}{\partial p_i}=0,
$
as it was discussed in Section \ref{section3}. (\ref{dane}) is the semiclassical manifestation of the chiral anomaly in any even dimensions.

\subsection{The Chiral Magnetic Effect}

Our formalism of kinetic theory directly provides the  solutions of the equations of motion (\ref{eomd}) for  $(\dot{X}_\ssA\tilde{W}_{1/2},
\tilde{W}_{1/2}\dot{P}_\ssA)$ as well as $\tilde{W}_{1/2}$ in terms of the phase space variables $(x_\ssA ,p_\ssA ),$ by equating the right hand sides of (\ref{led}) and (\ref{ln1}):
$$
\frac{(-1)^{n+1}}{(2n+1)!}d\tilde{W}_\ssH^{2n+1}=\frac{\partial}{\partial t} \tilde{W}_\ssh+\frac{\partial}{\partial x_\ssA}(\dot{X}_\ssA\tilde{W}_\ssh )+\frac{\partial}{\partial p_\ssA} (\tilde{W}_\ssh \dot{P}_\ssA).
$$
 To derive the chiral current it is sufficient to present $\dot{X}_\ssA\tilde{W}_{1/2}$ which  includes several terms like
\begin{eqnarray}
\dot{X}_\ssA\tilde{W}_{1/2}&=&\frac{(-1)^n}{(2n)!dVdt}dx_\ssA\w \hat{p}_\ssB dp_\ssB\w dt\w \overbrace{dp_\ssC\w dx_\ssC...dp_\ssD\w dx_\ssD}^{2n\ times} \nonumber \\
& &+\frac{(-1)^n}{(2n-1)! dVdt}dx_\ssA \w {\cal{E}}_\ssB dx_\ssB\w dt \w {\cal{G}}\overbrace{dp_\ssC\w dx_\ssC...dp_\ssD\w dx_\ssD}^{2n-1\ times}\nonumber  \\
& &+\frac{(-1)^{n+1}}{(2n-2)! dVdt}dx_\ssA\w \hat{p}_\ssB dp_\ssB\w dt \w {\cal{G}}{\cal{F}}\overbrace{dp_\ssC\w dx_\ssC...dp_\ssD\w dx_\ssD}^{2n-2\ times} \nonumber \\
& & +\ ...\ +\frac{1}{(n!)^2\ 2^{2n}}\epsilon_{\ssA\ssB\ssC...\ssD\ssE}\overbrace{{\cal{F}}_{\ssB\ssC}...{\cal{F}}_{\ssD\ssE}}^{n\ times}\epsilon_{\ssI\ssJ\ssK...\ssL\ssM}\hat{p}_\ssI\overbrace{{\cal{F}}_{\ssJ\ssK}...{\cal{F}}_{\ssL\ssM}}^{n\ times}. \label{geqm}
\end{eqnarray}

In the semiclassical approximation we define the current as
$$
j_\ssA=\int{\frac{d^{2n+1}p}{(2\pi)^{2n+1}}\Tr[\dot{X}_\ssA\tilde{W}_{1/2}]f(x,p,t)}.
$$
We deal with the CME which is generated by the terms depending on the external magnetic field ${\cal{F}}_{\ssA\ssB}.$ 
As it is shown in Appendix B, once we take 
the trace over the spin indices there remains only 
one term depending on the external magnetic field ${\cal{F}}_{\ssA\ssB},$
which is generated by the last term given in (\ref{geqm}). 

Therefore  the chiral magnetic current is calculated to be
\begin{eqnarray}
j_\ssA^{\ssC\ssM\ssE}
&=&\frac{1}{2^{2n}(n!)^2}\int{\frac{d^{2n+1}p}{(2\pi)^{2n+1}}\epsilon_{\ssA\ssB\ssC...\ssD\ssE}\overbrace{{\cal{F}}_{\ssB\ssC}...{\cal{F}}_{\ssD\ssE}}^{n\ times}\Tr[\epsilon_{\ssI\ssJ\ssK...\ssL\ssM}\hat{p}_\ssI\overbrace{{\cal{G}}_{\ssJ\ssK}...{\cal{G}}_{\ssL\ssM}}^{n\ times}]f(x,p,t)} \nonumber\\
&=&\frac{(-1)^{n+1}(2n)!}{2^{2n}(n!)^2}\int{\frac{d^{2n+1}p}{(2\pi)^{2n+1}}\epsilon_{\ssA\ssB\ssC...\ssD\ssE}\overbrace{{\cal{F}}_{\ssB\ssC}...{\cal{F}}_{\ssD\ssE}}^{n\ times}(\bm{\hat{p}}\cdot\bm{b})f(x,p,t)}, \label{cmecd}
\end{eqnarray}
where $\bm{b}$ is the Dirac monopole field given in (\ref{dmd}).
When we deal with an isotropic momentum distribution $f=f(E),$ the  angular part of (\ref{cmecd})  can be computed, so that we establish
the chiral magnetic current as
\begin{eqnarray*}
j_\ssA^{\ssC\ssM\ssE}&=&\frac{(-1)^{n+1}(2n)!}{2^{2n}(n!)^2}\frac{{\rm{Vol}}(S^{2n})}{2(2\pi)^{2n+1}}\epsilon_{\ssA\ssB\ssC...\ssD\ssE}\overbrace{{\cal{F}}_{\ssB\ssC}...{\cal{F}}_{\ssD\ssE}}^{n\ times}\int{dEf(E)},\\
&=&\frac{(-1)^{n+1}}{2^{n}(2\pi )^{n+1}n!}\epsilon_{\ssA\ssB\ssC...\ssD\ssE}\overbrace{{\cal{F}}_{\ssB\ssC}...{\cal{F}}_{\ssD\ssE}}^{n\ times}\int{dEf(E)}.
\end{eqnarray*} 
This is the chiral magnetic current conjectured in \cite{ls}. 

\subsection{The Chiral Vortical Effect}

When one considers the  vortical flow of the fluid in $3+1$ dimensions, there exists an induced current linear in the vorticity  which is 
known as the chiral vortical effect \cite{CVE, jerd, nbanar}. It is similar to the CME although the vorticity is an 
 intrinsic property of the fluid in contrast to the external magnetic field: The vorticity  $\omega_i\equiv (1/2)\epsilon_{ijk} \omega^{jk},$
 gives rise to the effective magnetic field $2|\bm{p}|\bm{\omega},$ where $|\bm{p}|$ is the energy of the co-moving Weyl particle \cite{KS11}. Hence, we can incorporate  $ \omega_{\ssA \ssB},$ denoting the vorticity in $d+1$ dimensions, into the present formalism by switching off the external fields and dealing with the symplectic two-form
$$
\tilde{W}_\ssH\equiv dp_\ssA\wedge dx_\ssA+\Omega-{\cal{G}}-\hat{p}_\ssA dp_\ssA\wedge dt,
$$
where $\Omega= |\bm{p}| \omega_{\ssA \ssB} dx_\ssA\w dx_\ssB .$  By performing the same computations which yielded the CME (\ref{cmecd}), we obtain the chiral vortical effect in any even dimensions for an isotropic momentum distribution as
$$
j_\ssA^{\ssC\ssV\ssE}=\frac{(-1)^{n+1}}{2^{n}(2\pi )^{n+1}n!}\epsilon_{\ssA\ssB\ssC...\ssD\ssE}\overbrace{\omega_{\ssB\ssC}...\omega_{\ssD\ssE}}^{n\ times}\int{dE\ E^nf(E)}.
$$
This is in accord with the chiral vortical effect conjectured in \cite{ls}.

\section{Discussions
\label{sectionD}}

We first calculated the Abelian Berry gauge field arising from the $3+1$ dimensional Weyl Hamiltonian and demonstrated  that  the Berry field strength  yields  a monopole situated at the origin of  phase space. In the Appendix, we explicitely showed that similar results hold in $5+1$ dimensions where the Berry gauge field is non-Abelian. In fact in any even $d+1$ dimensions the related Berry field strength engenders a Dirac monopole field. This monopole field is responsible for  the chiral anomaly manifested itself  in the kinetic theory of electrons as non-conservation of particle current. We showed that this monopole  is also the source of the CME. We presented an efficient method of finding the path integral measure and  solutions of the equations of motion for the first derivatives of the phase space variables weighted by this measure. 
This furnished the possibility of obtaining the chiral current directly inspecting  Liouville equation. 
Hence we calculated the CME and chiral anomaly in any even dimensions within the same formulation.
The accomplished CME and chiral vortical effect are in accord with the ones conjectured  in \cite{ls}. The main novelty  is to keep the spin dependence explicit without attributing them some dynamical variables. 

It seems that there is no obstacle of incorporating external  non-Abelian gauge fields to our method in the line with \cite{ds,ds1}.

A semiclassical study of   the massive Dirac particle engaging the non-Abelian Berry gauge fields was presented in \cite{mwp}. There 
some dynamical variables have been associated with the spin degrees of freedom.
In principle the approach of dealing with  the non-Abelian Berry gauge fields presented here  can  be adopted to perform a similar study 
in the ordinary classical phase space without enlarging it with some new dynamical variables.

The formalism based on the matrix valued symplectic form is not restricted to even dimensions. It can be employed to establish  solution of the equations of motion in terms of phase space variables in any dimensions. These solutions which exhibit the spin dependence explicitly can be
useful to formulate some interesting  physical phenomena  like the spin Hall effect.
\vspace{0.5cm}

\noindent
{\bf{ Acknowledgments}}\\
M.E. acknowledges the support from the Major State Basic Research Development Program in China (Grant No. 2015CB856903).

\vspace{1cm}
\noindent
{\bf\Large{Appendices}}

\appendix

\renewcommand{\theequation}{\thesection.\arabic{equation}}
\setcounter{equation}{0}

\section{  Chiral Kinetic Theory  in $5+1$ Dimensions
\label{section4}}

To illustrate the whole machinery explicitly we would like to discuss $d=5$ case which the lowest dimensions  where the Berry gauge fields are non-Abelian. A representation of $\alpha_i;\  i=1,..,5,$  satisfying (\ref{alf})
can be given as the direct product of the $3+1$ dimensional gamma matrices (\ref{cb}) with
$
\sigma_0={\rm diag } (1,1)
$ 
and $
\sigma_3={\rm diag } (1,-1):
$
$$
\alpha_1=\sigma_0\otimes\gamma_0\gamma_1, \quad \alpha_2=\sigma_0\otimes\gamma_0\gamma_2, \quad \alpha_3=\sigma_0\otimes\gamma_0\gamma_3,\quad \alpha_4=i\sigma_0\otimes\gamma_0\gamma_5, \quad \alpha_5=\sigma_3\otimes\gamma_0.
$$
In this representation  the $\bm{\Sigma}$ matrices are
$$
\Sigma_{a}=
\begin{pmatrix}
\sigma_{a} & 0\\
0 & -\sigma_{a}
\end{pmatrix};\ {\scriptstyle a=1,2,3},\quad
\Sigma_4=
\begin{pmatrix}
0 & i\\
-i & 0
\end{pmatrix}, \quad
\Sigma_5=
\begin{pmatrix}
0 & -1\\
-1 & 0
\end{pmatrix}.
$$
 Therefore,  the Weyl Hamiltonian is expressed as
$$
{\cal{H}}_\ssW=
\begin{pmatrix}
\sigma_ap_a & i(p_4+ip_5)\\
-i(p_4-ip_5) & -\sigma_ap_a 
\end{pmatrix}.
$$

We construct the
normalized, positive energy eigenstates  ${\cal{H}_\ssW} |\psi^{\alpha}\rangle=p|\psi^{\alpha}\rangle ;$ $\alpha =1,2,$ as
$$
|\psi^1\rangle=\sqrt{\frac{p_4^2+p_5^2}{2p(p-p_3)}}
\begin{pmatrix} 
\frac{i(p_1-ip_2)}{p_4-ip_5} \\  \frac{i(p-p_3)}{p_4-ip_5}  \\ 0 \\ 1 
\end{pmatrix}, \quad 
|\psi^2\rangle=\frac{1}{\sqrt{2p(p-p_3)}}
\begin{pmatrix} 
i(p_4+ip_5) \\ 0 \\ p-p_3 \\ -(p_1+ip_2)  
\end{pmatrix}.
$$
Plug these  degenerate eigenvectors into the definition (\ref{bgf}) to build the non-Abelian  Berry gauge field components 
 ${\cal{A}}_i^{\alpha\beta}$ as
\begin{eqnarray}
\label{ebg}
{\cal{A}}_1&=&\frac{1}{2p(p-p_3)}\begin{pmatrix}-p_2 & -i\sqrt{p_4^2+p_5^2} \\ i\sqrt{p_4^2+p_5^2} & p_2  \end{pmatrix}, \nonumber\\
{\cal{A}}_2&=&\frac{1}{2p(p-p_3)}\begin{pmatrix}p_1 & \sqrt{p_4^2+p_5^2} \\ \sqrt{p_4^2+p_5^2}  & -p_1  \end{pmatrix}, \nonumber \\
{\cal{A}}_3&=&0,\\
{\cal{A}}_4&=&\frac{1}{2p(p-p_3)}\begin{pmatrix}p_5[\frac{2p(p-p_3)}{p_4^2+p_5^2}-1] & \frac{i\sqrt{p_4^2+p_5^2} (p_1+ip_2)}{p_4+ip_5}\\ 
\frac{-i\sqrt{p_4^2+p_5^2} (p_1-ip_2)}{p_4-ip_5} & p_5  \end{pmatrix},\nonumber\\
{\cal{A}}_5&=&\frac{1}{2p(p-p_3)}\begin{pmatrix}-p_4[\frac{2p(p-p_3)}{p_4^2+p_5^2}-1] & \frac{-\sqrt{p_4^2+p_5^2} (p_1+ip_2)}{p_4+ip_5}\\ 
\frac{-\sqrt{p_4^2+p_5^2} (p_1-ip_2)}{p_4-ip_5} & -p_4  \end{pmatrix}. \nonumber
\end{eqnarray}
By employing (\ref{ebg}) in the definition (\ref{dbc}) we extract ${\cal{G}}_{ij}^{\alpha\beta}$ as follows,
\begin{eqnarray*}
{\cal{G}}_{12}&=&\scriptstyle \frac{1}{2p^3(p-p_3)}
\begin{pmatrix}
\scriptstyle (p_3^2+p_4^2+p_5^2)-pp_3 & \scriptstyle -\scriptstyle (p_1+ip_2)\sqrt{(p_4^2+p_5^2)}\\
 \scriptstyle -\scriptstyle (p_1-ip_2)\sqrt{(p_4^2+p_5^2)} & \scriptstyle -\scriptstyle (p_3^2+p_4^2+p_5^2)+pp_3 
\end{pmatrix},\nonumber \\
{\cal{G}}_{13}&=&\scriptstyle \frac{1}{2p^3}
\begin{pmatrix}
\scriptstyle p_2 & \scriptstyle i \scriptstyle \sqrt{(p_4^2+p_5^2)}\\
 \scriptstyle -\scriptstyle i \scriptstyle \sqrt{(p_4^2+p_5^2)}& \scriptstyle- \scriptstyle p_2
\end{pmatrix},\nonumber \\
{\cal{G}}_{14}&=&\scriptstyle \frac{1}{2p^3(p-p_3)}
\begin{pmatrix}
 \scriptstyle{p_1p_5-p_2p_4} & \scriptstyle - \scriptstyle \frac{(ip_4+p_5)\left(i(p_1p_2+p_4p_5)+p_3p-p_2^2-p_3^2-p_5^2\right)}{\sqrt{(p_4^2+p_5^2)}}\\
\scriptstyle \frac{(-ip_4+p_5)\left(i(p_1p_2+p_4p_5)-p_3p+p_2^2+p_3^2+p_5^2 \right)}{\sqrt{(p_4^2+p_5^2)}} & \scriptstyle p_2p_4-p_1p_5
\end{pmatrix},\nonumber \\
{\cal{G}}_{15}&=& \scriptstyle \frac{1}{2p^3(p-p_3)}
\begin{pmatrix}
\scriptstyle - \scriptstyle p_1p_4-p_2p_5 & \scriptstyle \frac{(p_4-ip_5)\left(i(p_1p_2-p_4p_5)+p_3p-p_2^2-p_3^2-p_4^2\right)}{\sqrt{(p_4^2+p_5^2)}}\\
\scriptstyle \frac{(p_4+ip_5)\left(-i(p_1p_2-p_4p_5)+p_3p-p_2^2-p_3^2-p_4^2\right)}{\sqrt{(p_4^2+p_5^2)}} & \scriptstyle p_1p_4+p_2p_5
\end{pmatrix},\nonumber \\
{\cal{G}}_{23}&=&\scriptstyle \frac{1}{2p^3}
\begin{pmatrix}
\scriptstyle - \scriptstyle p_1 & \scriptstyle - \scriptstyle \sqrt{(p_4^2+p_5^2)}\\
 \scriptstyle -\scriptstyle \sqrt{(p_4^2+p_5^2)}& \scriptstyle p_1
\end{pmatrix},\nonumber \\
{\cal{G}}_{24}&=&\scriptstyle \frac{1}{2p^3(p-p_3)}
\begin{pmatrix}
\scriptstyle p_1p_4+p_2p_5 & \scriptstyle \frac{(p_4-ip_5)\left(i(p_4p_5-p_1p_2)+p_3p-p_1^2-p_3^2-p_5^2\right)}{\sqrt{(p_4^2+p_5^2)}}\\
\scriptstyle \frac{(p_4+ip_5)\left(-i(p_4p_5-p_1p_2)+p_3p-p_1^2-p_3^2-p_5^2\right)}{\sqrt{(p_4^2+p_5^2)}} & \scriptstyle - \scriptstyle p_1p_4-p_2p_5
\end{pmatrix},\nonumber \\
{\cal{G}}_{25}&=&\scriptstyle \frac{1}{2p^3(p-p_3)}
\begin{pmatrix}
\scriptstyle p_1p_5-p_2p_4 & \scriptstyle \frac{(ip_4+p_5)\left(-i(p_4p_5+p_1p_2)+p_3p-p_1^2-p_3^2-p_4^2\right)}{\sqrt{(p_4^2+p_5^2)}}\\
\scriptstyle \frac{(-ip_4+p_5)\left(i(p_4p_5+p_1p_2)+p_3p-p_1^2-p_3^2-p_4^2\right)}{\sqrt{(p_4^2+p_5^2)}}& \scriptstyle - \scriptstyle p_1p_5+p_2p_4
\end{pmatrix},\nonumber \\
{\cal{G}}_{34}&=&\scriptstyle \frac{1}{2p^3}
\begin{pmatrix}

\scriptstyle - \scriptstyle p_5 & \scriptstyle \frac{i(p_1+ip_2)\sqrt{(p_4^2+p_5^2)}}{p_4+ip_5}\\
\scriptstyle \frac{-i(p_1-ip_2)\sqrt{(p_4^2+p_5^2)}}{p_4-ip_5}& \scriptstyle p_5
\end{pmatrix},\nonumber \\
{\cal{G}}_{35}&=&\scriptstyle \frac{1}{2p^3}
\begin{pmatrix}
\scriptstyle p_4 & \scriptstyle \frac{-(p_1+ip_2)\sqrt{(p_4^2+p_5^2)}}{p_4+ip_5}\\
\scriptstyle \frac{-(p_1-ip_2)\sqrt{(p_4^2+p_5^2)}}{p_4-ip_5}& \scriptstyle - \scriptstyle p_4
\end{pmatrix},\nonumber \\
{\cal{G}}_{45}&=&\scriptstyle \frac{1}{2p^3(p-p_3)}
\begin{pmatrix}
\scriptstyle (p_1^2+p_2^2+p_3^2)-pp_3 & \scriptstyle (p_1+ip_2)\sqrt{(p_4^2+p_5^2)}\\
 \scriptstyle (p_1-ip_2)\sqrt{(p_4^2+p_5^2)} & \scriptstyle - \scriptstyle (p_1^2+p_2^2+p_3^2)+pp_3 
\end{pmatrix}.\nonumber \\
\end{eqnarray*}
They are all traceless.

The symplectic two-form, the Pfaffian and the matrix valued vector field  are denoted, respectively,   $\tilde{w}_\ssH ,$ 
 $\tilde{w}_\ssH ,$  and  $\tilde{v} .$
We expose $\tilde{\Omega}$ explicitly as
\begin{eqnarray}
\tilde{w}_\ssH^5 &=& dp_i\w dx_i\w dp_j\w dx_j\w dp_k\w dx_k\w dp_l\w dx_l\w dp_m\w dx_m \nonumber \\
&-&20{\cal{F}}\w {\cal{G}}\w dp_i\w dx_i\w dp_j\w dx_j\w dp_k\w dx_k+30{\cal{F}}\w {\cal{F}}\w {\cal{G}}\w {\cal{G}}\w dp_i\w dx_i\nonumber\\
&-&5\hat{p}_idp_i\w dt\w  dp_j\w dx_j\w dp_k\w dx_k\w dp_l\w dx_l\w dp_m\w dx_m\nonumber\\
&-&20{\cal{E}}_i{\cal{G}}\w dx_i\w dt\w dp_j\w dx_j\w dp_k\w dx_k\w dp_l\w dx_l\nonumber\\
&+&60\hat{p}_i{\cal{G}}\w {\cal{F}}\w dp_i\w dt\w dp_j\w dx_j\w dp_k\w dx_k\label{w5} \\
&+&60{\cal{E}}_i{\cal{G}}\w {\cal{G}}\w {\cal{F}}\w dx_i\w dt\w dp_j\w dx_j-30\hat{p}_i{\cal{F}}\w {\cal{F}}\w {\cal{G}}\w {\cal{G}}\w dp_i\w dt \nonumber\\
&+&5{\cal{E}}_idx_i\w dt\w  dp_j\w dx_j\w dp_k\w dx_k\w dp_l\w dx_l\w dp_m\w dx_m\nonumber\\
&-&20\hat{p}_i{\cal{F}}\w dp_i\w dt\w  dp_j\w dx_j\w dp_k\w dx_k\w dp_l\w dx_l\nonumber\\
&-&60{\cal{E}}_i{\cal{F}}\w {\cal{G}}\w dx_i\w dt\w dp_j\w dx_j\w dp_k\w dx_k\nonumber\\
&+&60\hat{p}_i{\cal{G}}\w {\cal{F}}\w {\cal{F}}\w  dp_i\w dt\w  dp_j\w dx_j+30 {\cal{E}}_i{\cal{G}}\w {\cal{G}}\w {\cal{F}}\w {\cal{F}}\w dx_i\w dt. \nonumber
\end{eqnarray}

Although it is cumbersome, making use of  $ {\cal{G}}_{ij}$ presented above we obtain
 \be
\label{aterm}
\Tr [L_{\tilde{v}}\tilde{\Omega}]=-\frac{\pi^2}{2}\delta^5(p)\epsilon^{ijklm}{\cal{E}}_i{\cal{F}}_{jk}{\cal{F}}_{lm}\ dV^{(5)}\w dt.
\ee
Hence, the chiral current is anomalous:
$$
\frac{\partial n}{\partial t}+\vec{\nabla} \cdot \vec{j}=-\frac{1}{ (4\pi )^3}f(x,p=0,t)\epsilon^{ijklm}{\cal{E}}_i{\cal{F}}_{jk}{\cal{F}}_{lm}.
$$

By inspecting  (\ref{w5}) and comparing with the formal expression (\ref{led}) for $d=5,$
we read directly:
\begin{eqnarray}
\label{equal5}
\tilde{w}_{1/2}&=&I+\frac{1}{2}{\cal{F}}_{ij}{\cal{G}}_{ij}+\frac{1}{8}{\cal{F}}_{ij}{\cal{F}}_{kl}({\cal{G}}_{ij}{\cal{G}}_{kl}+{\cal{G}}_{ik}{\cal{G}}_{lj}+{\cal{G}}_{lj}{\cal{G}}_{ik})\nonumber\\
\dot{X}_r\tilde{w}_{1/2}&=&\hat{p}_r+{\cal{E}}_i{\cal{G}}_{ri}+\frac{1}{2}{\cal{F}}_{ij}(\hat{p}_r{\cal{G}}_{ij}+2\hat{p}_j{\cal{G}}_{ri})+\nonumber\\
&&\frac{1}{4}{\cal{E}}_i{\cal{F}}_{jk}({\cal{G}}_{ri}{\cal{G}}_{jk}+{\cal{G}}_{jk}{\cal{G}}_{ri}+2{\cal{G}}_{rj}{\cal{G}}_{ki}+2{\cal{G}}_{ki}{\cal{G}}_{rj})+\frac{1}{64}{\cal{F}}_{ij}{\cal{F}}_{kl}{\cal{G}}_{mn}{\cal{G}}_{st}\hat{p}_p\epsilon_{rijkl}\epsilon_{mnstp}\\
\tilde{w}_{1/2}\dot{P}_r&=&{\cal{E}}_r+\hat{p}_i{\cal{F}}_{ri}+\frac{1}{2}{\cal{G}}_{ij}({\cal{E}}_r{\cal{F}}_{ij}+2{\cal{E}}_j{\cal{F}}_{ri})+\frac{1}{2}\hat{p}_i{\cal{G}}_{jk}({\cal{F}}_{ri}{\cal{F}}_{jk}+2{\cal{F}}_{rj}{\cal{F}}_{ki})\nonumber\\
&&+\frac{1}{64}{\cal{F}}_{mn}{\cal{F}}_{st}{\cal{E}}_p{\cal{G}}_{ij}{\cal{G}}_{kl}\epsilon_{rijkl}\epsilon_{mnstp} .\nonumber
\end{eqnarray}
Now, it is possible to build the the classical current by means of the phase space probability function $f(x,p,t)$ and taking  the trace  over the spin indices of the solution
(\ref{equal5}) as
\begin{eqnarray}
\label{cme5}
j_r &=&\int \frac{d^5p}{(2\pi)^5}\Tr[\dot{X}_r\tilde{w}_{1/2}] f\nonumber\\
&=&2\int \frac{d^5p}{(2\pi)^5}\hat{p}_rf+\frac{1}{2}\int \frac{d^5p}{(2\pi)^5}{\cal{E}}_i{\cal{F}}_{jk}\Tr[{\cal{G}}_{ri}{\cal{G}}_{jk}+2{\cal{G}}_{rj}{\cal{G}}_{ki}]f\\
&&+\frac{1}{64}\int \frac{d^5p}{(2\pi)^5}{\cal{F}}_{ij}{\cal{F}}_{kl}\epsilon_{rijkl}\Tr[{\cal{G}}_{mn}{\cal{G}}_{st}\hat{p}_p\epsilon_{mnstp}]f.\nonumber
\end{eqnarray}
The last term in (\ref{cme5}) gives  the $5+1$ dimensional chiral magnetic effect 
$$
j_r^{\scriptscriptstyle{CME}}=-\frac{3}{8}\int \frac{d^5p}{(2\pi)^5}{\cal{F}}_{ij}{\cal{F}}_{kl}\epsilon_{rijkl}(\bm{\hat{p}}\cdot\bm{b})f.
$$
When the Fermi-Dirac distribution is considered, it yields 
$j_r^{\scriptscriptstyle{CME}} =(-\mu /8(2\pi)^3) \epsilon_{rijkl} {\cal{F}}_{ij}{\cal{F}}_{kl},$  at finite chemical potential $\mu . $

\renewcommand{\theequation}{\thesection.\arabic{equation}}
\setcounter{equation}{0}

\section{Properties of the Berry Curvature and the $\Sigma$ Matrices}

 Interrelation between the Dirac monopole and the Berry curvature was established 
 for an even $d+1$ dimensional Weyl Hamiltonian in \cite{me}. Here, we  review the arguments of \cite{me} and also demonstrate the 
trace properties of the wedge products of the Berry curvature ${\cal{G}}$  used in Section \ref{section5}.

In terms of the energy eigenstates one can introduce a unitary matrix $U$ which diagonalizes the Weyl Hamiltonian (\ref{WHa}),
$$
U{\cal{H}}_{\ssW} U^\dagger=p({\cal{I}}^+-{\cal{I}}^-).
$$
${\cal{I}}^+$ and ${\cal{I}}^-$ project onto the positive and negative energy subspaces. The Berry gauge field (\ref{bgf}) can be written by means of $U$ as 
$$
{\cal{\bm{A}}}=i{\cal{I}}^+U\partial_{\bm{p}}U^\dagger {\cal{I}}^+.
$$
Thus, one can construct the Berry curvature (\ref{dbc}) in terms of $U$. 

In $d=2n+1$ dimensions the trace of the $m\leq n$ subsequent Berry field strengths can be expressed as  
\be
\label{app1}
\epsilon_{\ssA_1 \ssA_2 \dots \ssA_{2m+1}\dots \ssA_{2n+1}}
 \Tr [{\cal{G}}_{\ssA_2\ssA_3}...{\cal{G}}_{\ssA_{2m}\ssA_{2m+1}}]=(2i)^m
\epsilon_{\ssA_1 \ssA_2 \dots \ssA_{2m+1}\dots \ssA_{2n+1}}\Tr[P^+(\partial_{\ssA_2} P^+)...(\partial_{\ssA_{2m+1}} P^+)].
\ee
We introduced
$$
\quad P^+=U^{\dagger} {\cal{I}}^+ U,
$$
which can be written as
$$
P^+=\frac{1}{2}(\frac{{\cal{H}}_\ssW}{p}+1).
$$ 
Plugging it into  (\ref{app1}) leads to 
\begin{eqnarray}
\label{app2}
\epsilon_{\ssA_1 \ssA_2 \dots \ssA_{2m+1}\dots \ssA_{2n+1}}\Tr[P^+(\partial_{\ssA_2} P^+)\dots (\partial_{\ssA_{2m+1}} P^+)]
&=&\frac{\epsilon_{\ssA_1 \ssA_2 \dots \ssA_{2m+1}\dots \ssA_{2n+1}}}{(2p)^{2m+1}}\Tr[{\cal{H}}_\ssW(\partial_{\ssA_2}{\cal{H}}_\ssW)\dots (\partial_{\ssA_{2m+1}}{\cal{H}}_\ssW)] \nonumber \\
&=&\frac{\epsilon_{\ssA_1 \ssA_2 \dots \ssA_{2m+1}\dots \ssA_{2n+1}}}{(2p)^{2m+1}}\Tr[\bm{\Sigma}\cdot\bm{p}\Sigma_{\ssA_2}\dots \Sigma_{\ssA_{2m+1}}]. 
\end{eqnarray}

Inspecting (\ref{app1}) and (\ref{app2}) one observes that the trace of the wedge products of the Berry curvature, ${\cal{G}},$ can be expressed in terms of the trace of the antisymmetrized $\Sigma$ matrices:
$$
\epsilon_{\ssA_1 \ssA_2 \dots \ssA_{2m+1}\dots \ssA_{2n+1}} \Tr [{\cal{G}}_{\ssA_2\ssA_3}\dots{\cal{G}}_{\ssA_{2m}\ssA_{2m+1}}]=(2i)^m
\frac{p_\ssA}{(2p)^{2m+1}}\epsilon_{\ssA_1 \ssA_2 \dots \ssA_{2m+1}\dots \ssA_{2n+1}}\Tr[\Sigma_\ssA\Sigma_{\ssA_2}\dots\Sigma_{\ssA_{2m+1}}].
$$

$\Sigma_{\ssA}$   obey the Clifford algebra,
$$
\{\Sigma_{\ssA},\Sigma_{\ssB}\}=2\delta_{\ssA\ssB}.
$$
Moreover, they are traceless,
$$
\Tr[\Sigma_\ssA]=0,
$$
and in $2n+2$ dimensional spacetime they satisfy the following identity,
$$
\Sigma_1...\Sigma_{2n+1}=i^{n+2}\bm{1}_{\scriptscriptstyle{2^n\times 2^n}}.
$$
Thus  the trace of $2n+1$ antisymmetric product of the $\Sigma$ matrices yields
$$
\frac{1}{(2n+1)!}\epsilon_{\ssA_1\dots\ssA_{2n+1}}\Tr[\Sigma_{\ssA_1}\dots\Sigma_{\ssA_{2n+1}}]=i^{n+2}2^n.
$$

Actually one can observe that the trace of the product of even number of different $\Sigma$ matrices always vanishes because of satisfying the Clifford algebra:
$$
\Tr[\Sigma_{\ssA_1}\dots\Sigma_{\ssA_{2m}}]=0.
$$
Moreover, it can be easily shown that the trace of the product of $2m+1$ different $\Sigma$ matrices is equal to the trace of the product of the remaining $2(n-m)$ $\Sigma$ matrices which is equal to zero. Therefore, the trace of the antisymmetrized product of the Berry field strength vanishes
$$
\epsilon^{\ssA_1\ssA_2\dots\ssA_{2m-1}\ssA_{2m}\dots\ssA_{2n+1}} \Tr [{\cal{G}}_{\ssA_1\ssA_2}\dots{\cal{G}}_{\ssA_{2m-1}\ssA_{2m}}]=0,
$$ 
for the case $m<n$. 
When $m=n$ one finds
$$
\epsilon_{\ssA_1 \ssA_2\ssA_3\dots\ssA_{2n}\ssA_{2n+1}} \Tr [{\cal{G}}_{\ssA_2\ssA_3}\dots{\cal{G}}_{\ssA_{2n}\ssA_{2n+1}}]=(-1)^{n+1}
(2n)!\frac{p_{\ssA_1}}{2p^{2n+1}},
$$ 
which is the Dirac monopole field.

\newcommand{\PRL}{Phys. Rev. Lett. }
\newcommand{\PRB}{Phys. Rev. B }
\newcommand{\PRD}{Phys. Rev. D }

\end{document}